\documentclass[preprint,showpacs,preprintnumbers,amsmath,amssymb,superscriptaddress]{revtex4}
\usepackage{graphicx}
\usepackage{dcolumn}
\usepackage{bm}

\begin{document}
\newcommand{\sgr}{SGR~1806-20 }

\title{Limits on the high-energy gamma and neutrino fluxes from the
SGR 1806-20 giant flare of December 27th, 2004 with the AMANDA-II
detector}

\affiliation{Dept.~of Physics and Astronomy, University of Alaska Anchorage, 3211 Providence Dr., Anchorage, AK 99508, USA}
\affiliation{CTSPS, Clark-Atlanta University, Atlanta, GA 30314, USA}
\affiliation{Dept.~of Physics, Southern University, Baton Rouge, LA 70813, USA}
\affiliation{Dept.~of Physics, University of California, Berkeley, CA 94720, USA}
\affiliation{Institut f\"ur Physik, Humboldt Universit\"at zu Berlin, D-12489 Berlin, Germany}
\affiliation{Lawrence Berkeley National Laboratory, Berkeley, CA 94720, USA}
\affiliation{Universit\'e Libre de Bruxelles, Science Faculty CP230, B-1050 Brussels, Belgium}
\affiliation{Vrije Universiteit Brussel, Dienst ELEM, B-1050 Brussels, Belgium}
\affiliation{Dept.~of Physics, Chiba University, Chiba 263-8522 Japan}
\affiliation{Dept.~of Physics and Astronomy, University of Canterbury, Private Bag 4800, Christchurch, New Zealand}
\affiliation{Dept.~of Physics, University of Maryland, College Park, MD 20742, USA}
\affiliation{Dept.~of Physics, Universit\"at Dortmund, D-44221 Dortmund, Germany}
\affiliation{Dept.~of Subatomic and Radiation Physics, University of Gent, B-9000 Gent, Belgium}
\affiliation{Max-Planck-Institut f\"ur Kernphysik, D-69177 Heidelberg, Germany}
\affiliation{Dept.~of Physics and Astronomy, University of California, Irvine, CA 92697, USA}
\affiliation{Dept.~of Physics and Astronomy, University of Kansas, Lawrence, KS 66045, USA}
\affiliation{Blackett Laboratory, Imperial College, London SW7 2BW, UK}
\affiliation{Dept.~of Astronomy, University of Wisconsin, Madison, WI 53706, USA}
\affiliation{Dept.~of Physics, University of Wisconsin, Madison, WI 53706, USA}
\affiliation{Institute of Physics, University of Mainz, Staudinger Weg 7, D-55099 Mainz, Germany}
\affiliation{University of Mons-Hainaut, 7000 Mons, Belgium}
\affiliation{Bartol Research Institute, University of Delaware, Newark, DE 19716, USA}
\affiliation{Dept.~of Physics, University of Oxford, 1 Keble Road, Oxford OX1 3NP, UK}
\affiliation{Institute for Advanced Study, Princeton, NJ 08540, USA}
\affiliation{Dept.~of Physics, University of Wisconsin, River Falls, WI 54022, USA}
\affiliation{Dept.~of Physics, Stockholm University, SE-10691 Stockholm, Sweden}
\affiliation{Dept.~of Astronomy and Astrophysics, Pennsylvania State University, University Park, PA 16802, USA}
\affiliation{Dept.~of Physics, Pennsylvania State University, University Park, PA 16802, USA}
\affiliation{Division of High Energy Physics, Uppsala University, S-75121 Uppsala, Sweden}
\affiliation{Dept.~of Physics and Astronomy, Utrecht University/SRON, NL-3584 CC Utrecht, The Netherlands}
\affiliation{Dept.~of Physics, University of Wuppertal, D-42119 Wuppertal, Germany}
\affiliation{DESY, D-15735 Zeuthen, Germany}

\author{A.~Achterberg}
\affiliation{Dept.~of Physics and Astronomy, Utrecht University/SRON, NL-3584 CC Utrecht, The Netherlands}
\author{M.~Ackermann}
\affiliation{DESY, D-15735 Zeuthen, Germany}
\author{J.~Adams}
\affiliation{Dept.~of Physics and Astronomy, University of Canterbury, Private Bag 4800, Christchurch, New Zealand}
\author{J.~Ahrens}
\affiliation{Institute of Physics, University of Mainz, Staudinger Weg 7, D-55099 Mainz, Germany}
\author{K.~Andeen}
\affiliation{Dept.~of Physics, University of Wisconsin, Madison, WI 53706, USA}
\author{D.~W.~Atlee}
\affiliation{Dept.~of Physics, Pennsylvania State University, University Park, PA 16802, USA}
\author{J.~N.~Bahcall}
\thanks{Deceased}
\affiliation{Institute for Advanced Study, Princeton, NJ 08540, USA}
\author{X.~Bai}
\affiliation{Bartol Research Institute, University of Delaware, Newark, DE 19716, USA}
\author{B.~Baret}
\affiliation{Vrije Universiteit Brussel, Dienst ELEM, B-1050 Brussels, Belgium}
\author{M.~Bartelt}
\affiliation{Dept.~of Physics, Universit\"at Dortmund, D-44221 Dortmund, Germany}
% ...

\date{\today}

\collaboration{IceCube Collaboration}
\noaffiliation

\begin{abstract}
On December 27th 2004, a giant $\gamma$-flare from the Soft Gamma-ray
Repeater 1806-20 saturated many satellite gamma-ray detectors. This
event was by more than two orders of magnitude the brightest cosmic
transient ever observed. If the gamma emission extends up to TeV
energies with a hard power law energy spectrum, photo-produced muons
could be observed in surface and underground arrays.  Moreover,
high-energy neutrinos could have been produced during the SGR giant
flare if there were substantial baryonic outflow from the magnetar.
These high-energy neutrinos would have also produced muons in an
underground array.  AMANDA-II was used to search for downgoing muons
indicative of high-energy gammas and/or neutrinos. The data revealed
no significant signal. The upper limit on the gamma flux at 90\% CL is
$dN/dE <$ 0.05 (0.5) TeV$^{-1}$~m$^{-2}$~s$^{-1}$ for $\gamma=-1.47$
($-2$).  Similarly, we set limits on the normalization constant of the
high-energy neutrino emission of $0.4$ ($6.1$) TeV$^{-1}$ m$^{-2}$
s$^{-1}$ for $\gamma=-1.47$ ($-2$).

\end{abstract}


\maketitle


{\it Introduction.---} Soft Gamma-ray Repeaters (SGRs) are X-ray pulsars which have quiescent
soft (2-10 keV) periodic X-ray emissions with periods ranging from 5
to 10~s and luminosities of the order of $10^{33-35}$~erg/s. They
exhibit repetitive bursts lasting $\sim 0.1$~s which reach peak
luminosities of $\sim 10^{41}$~erg/s in X-rays and
$\gamma$-rays. There are four known SGRs, three in the Milky Way
(including SGR 1860-20) and one in the Large Magellanic Cloud.  Three
of the four known SGRs have had hard spectrum ($\sim$MeV energy) giant
flares with luminosities reaching up to $\sim 10^{47}$ erg/s.  The
first of these giant flares (from SGR 0525-66~\cite{sgr1}) was
observed on March 5, 1979 by the Venera~11 and 12
spacecraft. SGR~1900+14 exhibited a giant flare in 1998~\cite{sgr1}.
The most recent and brightest flare came from \sgr on Dec. 27,
2004. This flare lasted about 5 minutes (the duration of the initial
spike was $\sim0.2$~s), had a peak luminosity of $\sim 2 \cdot
10^{47}$~erg/s and a total energy emission of $\sim 5 \cdot
10^{46}$~erg~\cite{Woods}.  This event was observed by several
satellite experiments~\cite{integral,swift-bat,rhessi}, although they
saturated during the blast. Recent estimates locate the source at a
distance of $15.1_{-1.3}^{+1.8}$~kpc~\cite{Corbel}, but this value is
still under debate~\cite{McClure-Griffiths}.

The favored ``magnetar" model for these objects is a neutron star with
a huge magnetic field ($B\sim10^{15}~G$). These giant flares can be
explained as global crustal fractures due to magnetic field
rearrangements liberating a high flux of X-rays and
$\gamma$-rays~\cite{Thompson}.

During the flare, the $\gamma$-ray spectrum up to $\sim$ 1 MeV is well
described with a blackbody spectrum with $kT = 175$~keV~\cite{rhessi},
though the last energy bin may leave room for a power-law
component. Similar considerations apply to the measurement up to
10~MeV in Ref.~\cite{Palmer}. However, fits to the data favor the presence of
a non-thermal component. Fits using a blackbody + power law (BB+PL)
spectrum show that the PL spectral index starts to harden months
before the giant flare up to values of about $-1.3 \div
-1.4$~\cite{Woods} below 10 keV.

The possibility of using underground detectors to observe the muons
produced in the electromagnetic showers induced by TeV gammas
generated in these flares was presented in Ref.~\cite{ourtheory}.
There have been suggestions of substantial baryonic outflow and the
possibility of high-energy neutrino production.  Radio
observations~\cite{Gaensler, Cameron} indicate an expanding radio
source with velocity 0.25-0.40$c$.  Gelfand {\em et
al.}~\cite{Gelfand} argue that a re-brightening of the radio emissions
$\sim$20 days after the giant flare can be explained if substantial amounts
of released energy went into a baryonic fireball, and make predictions
for TeV neutrino production.  Ioka {\sl et al.}~\cite{Ioka} also argue
that high-energy neutrino production can be related to the fraction of
burst energy released in the form of baryons.

The Dec. 2004 giant flare represents an excellent opportunity to probe
the high-energy spectrum of these sources by looking for events
correlated in time and space with this flare.  In this paper we
present the results of a search for a gamma and/or neutrino signal
during the \sgr giant flare using data from the AMANDA-II detector.
The short duration of these events and the fact that they come from a
point-like source result in negligible atmospheric muon and neutrino
backgrounds.  For the first time in AMANDA (see~\cite{macro} for an
example of previous search in other kind of sources), we used
down-going muons to look for muon photo-production in the atmosphere,
already proposed in Ref.~\cite{halzen}.  A search for coincidences
using gravitational waves was presented in Ref.~\cite{auriga}.

AMANDA (Antarctic Muon And Neutrino Detector Array) is currently
running in its AMANDA-II configuration of a 3D array of 677 optical
modules (OMs) distributed along 19 strings deployed at depths of
1500-2000~m in the South Pole ice~\cite{andres}.  These 8-inch
photomultipliers, enclosed in pressure-resistant glass spheres, make
it possible to reconstruct direction and energy of relativistic muons
through timing and intensity of the Cherenkov light. The ice layer
above the detector reduces the background of atmopheric muons by more
than 5 orders of magnitude compared to the surface flux.  Events are
recorded when at least 24 OMs register a signal within 2 $\mu$sec.
The detector rate on December 27th was 90~Hz (close to the AMANDA-II average).


{\it Analysis technique.---} For events like the burst of the \sgr in Dec. 2004, in which most of
the energy emitted during the flare is concentrated in a 1~s time
scale, the precise time and location of the event imply that the
background of atmotspheric muons becomes negligible~\cite{ourtheory},
thereby allowing a search for TeV $\gamma$-rays and downgoing TeV
neutrinos.

The IceCube collaboration follows a policy of blindness in its
analysis strategies. By studying the expected backgrounds and signals
prior to looking at the data, the analysis can be designed in an
unbiased fashion. In the case of expected small signals, this is
particularly relevant for having a clear procedure to determine the
probability of an event to be produced by background.  Thus, in this
analysis, the determination of the optimum selection criteria is done
using the simulation of the signal and comparing it with the expected
background.  The procedure, similar to that used in the search for
upward moving neutrinos from gamma-ray bursts~\cite{grbs}, is as follows:

\begin{itemize}

\item the background on-source and off-time is calculated using real
data, keeping blind 10 min around the burst onset. About one day of
off-time data was used to monitor the stability of the detector;
\item the signal from the source is simulated in order
to estimate the angular resolution and the effective area of the
detector;
\item the appropriate time window is estimated, based on the flare
onset times given by different X-ray satellites and their counting
rates;
\item the optimum search bin size is found by minimizing the Model
Discovery Factor (MDF)~\cite{hill}, defined as

   \begin{equation}
     \label{eq:mdf}
     MDF=\frac{\mu(n_b, CL, SP)}{n_s}
   \end{equation}

\noindent where $\mu$ is the Poisson mean of the number of signal
   events which would result in rejection of the background
   hypothesis, at the chosen confidence level $CL$, in $SP$\% of
   equivalent measurements. $SP$ stands for statistical power and
   $n_s$ is the number of signal events predicted by the model.
This definition is analogous to the Model Rejection Factor (MRF see
Ref.~\cite{hill}) that is used for setting upper limits. In the case
of the MDF, the bin size is optimized to maximize the probability of
discovery (for CL corresponding to five sigma and SP=90\%);

\item once the optimum search bin size has been found, the unblinding
of the data is done, i.e. the events inside the time window and the
search bin are counted;

\item this number of events is translated into a flux or a flux limit
if no significant excess on top of the background is found through
the knowledge of the expected signal in the detector for the given
analysis cuts.

\end{itemize}

Although both TeV $\gamma$-rays and neutrinos produce muons in the
detector array, the optimal choice of selection criteria depends on the assumed
signal.  The analysis was optimized to the TeV $\gamma$ signal.  Any
further optimization to the neutrino signal would be more than offset
by the penalty for an additional trials factor.


{\it Data and simulation.---} In order to have a background estimate for the flare, a time and
angular window have to be defined around the flare of equatorial
coordinates (J2000) right ascension = 18h 08m 39.34s and declination =
$-20^{\circ}$ 24' 39.7"~\cite{Cameron, Kaplan, Hurley}. The bulk of
the flare energy was concentrated in less than 0.6~s.  The Swift-BAT
counting rate drops by more than 2 orders of magnitudes after 0.6~s
from the onset of the burst~\cite{Palmer}.  Based on the observation
time for each
satellite~\cite{geotail,private,integral,auriga,rhessi,cluster4}, and
accounting for its position, the expected signal time in AMANDA was
calculated. The spread of the resulting times indicate that a safe
window is 1.5~s, centered at 21h 30m 26.6s of Dec. 27th, the onset
time of the flare.

To evaluate the performance of the detector, simulations were
performed for several input signals. The
CORSIKA-QGSJet01~\cite{corsika} and ANIS~\cite{anis} codes were used
to simulate the photon (and proton) and neutrino interactions,
respectively. The generated energy for photons is 10~TeV to
$10^{5}$~TeV and for protons is 10~GeV to $10^{5}$~TeV. The muons were
propagated in the ice with MMC~\cite{mmc} and the program
AMASIM~\cite{amasim} simulated the response of the detector. The
tracks were reconstructed with the same iterative log-likelihood
fitting procedure that was applied to the real data.

\begin{figure}
 \begin{center}
\includegraphics[width=1.0\linewidth,angle=0]{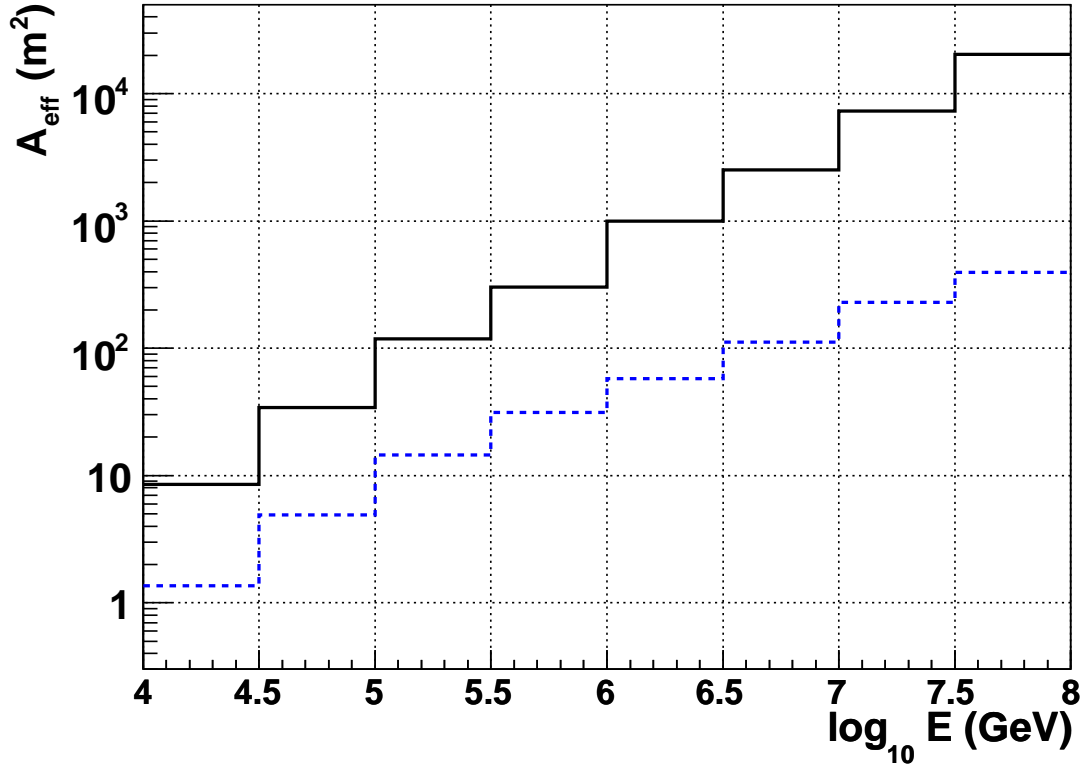} \\
  \end{center}
\caption{Effective area after reconstruction and track quality
 selection for gammas (solid, black) and neutrinos (blue, dashed).}
 \label{fig:effareas}
\end{figure}

The angular resolution for different cuts has been studied using the
simulation of down-going muons generated by cosmic rays. The angular
resolution, defined as the median of the angular difference between
the true and the reconstructed track, is $3.5^{\circ}$. This value was
obtained using atmospheric down-going muon high statistics simulations
and it was checked that this result is robust within 0.1~deg for the
expected signals of photons and neutrino induced muons. We also
considered a variety of spectral indices for the $\gamma$-ray spectrum
assuming values given in Ref.~\cite{ourtheory} and we found that the
angular resolution is almost independent of the spectral index.
The effective areas, defined as the equivalent area for a perfect
detector that is able to detect particles with 100\% efficiency, for
gammas and neutrinos are shown in figure~\ref{fig:effareas} as a
function of the energy.

\begin{figure}
 \begin{center}
\includegraphics[width=1.0\linewidth,angle=0]{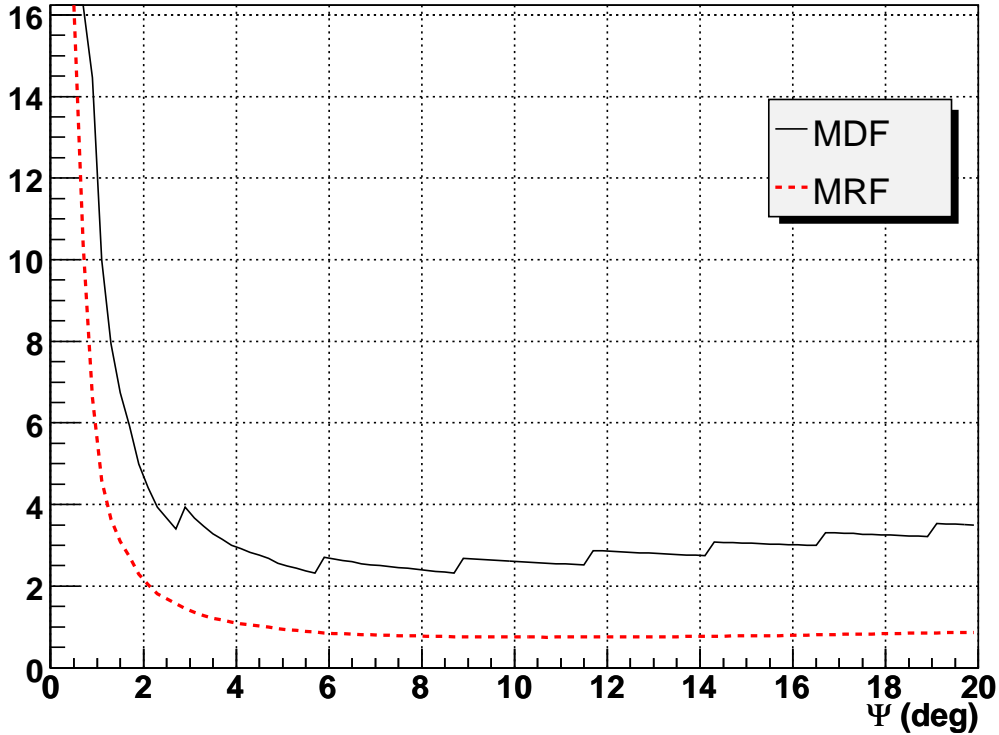}
 \end{center}
 \caption{Model discovery factor (solid, black) and model rejection
 factor (dashed, red) for an $E^{-1.47}$ spectrum.}
 \label{fig:mdf-mrf}
\end{figure}

As noted earlier, the optimum angular window is
determined by minimizing the Model discovery factor. In
figure~\ref{fig:mdf-mrf} we show the dependence of the MDF and MRF on
the circular angular window around the source position. The steps in
the MDF curve are due to the discreteness of the Poisson distribution. It
can also be seen that the MRF minimum interval is quite broad, indicating that
this variable is not very sensitive to the increase in the number of
background events with the increase of the angular window. This is due
to the small value of the background in the allowed time window.

The search bin size which optimizes the probability of discovery is
at 5.8$^{\circ}$. With this cut and a 1.5~s time window, the expected
background is 0.06 counts.  For events that satisfy the detector
trigger, we keep almost 80\% of the signal in this angular window.


{\it Systematic uncertainties.---} Several sources of systematic
uncertainties have been considered in this analysis. The uncertainty
of the hadronic model calculation is estimated to about 15\%, mostly
due to the unknown fraction of diffractive $\rho$
production~\cite{stanev}.  We have compared the muon yields using two
different models -CORSIKA/QGSJET and an analytic
calculation~\cite{halzen}- and found that they agreed to within
5\%. The uncertainty in the detector efficiency (20\%) comes mainly
from the overall sensitivity of the OMs and the optical properties of
the ice. The effect has been estimated simulating different reasonable
ice models and OM sensitivities.

The stability of the run was checked in order to exclude possible
non-particle events induced by detector electronics. These events are
identified by a specific method~\cite{arvid} looking for anomalous
values in a set of defined variables. A correction is made for the
electronics dead time (17\%, which is a typical value in normal runs).
Finally, the simulated and measured distributions of an extensive set
of variables, like zenith angle, number of hit optical modules, and
hit times, were compared in the search of possible anomalies. In all
the cases the agreement was within the systematic errors estimated
above.

\begin{figure}
 \begin{center}
\begin{tabular}{cc}
\includegraphics[width=1.0\linewidth,angle=0]{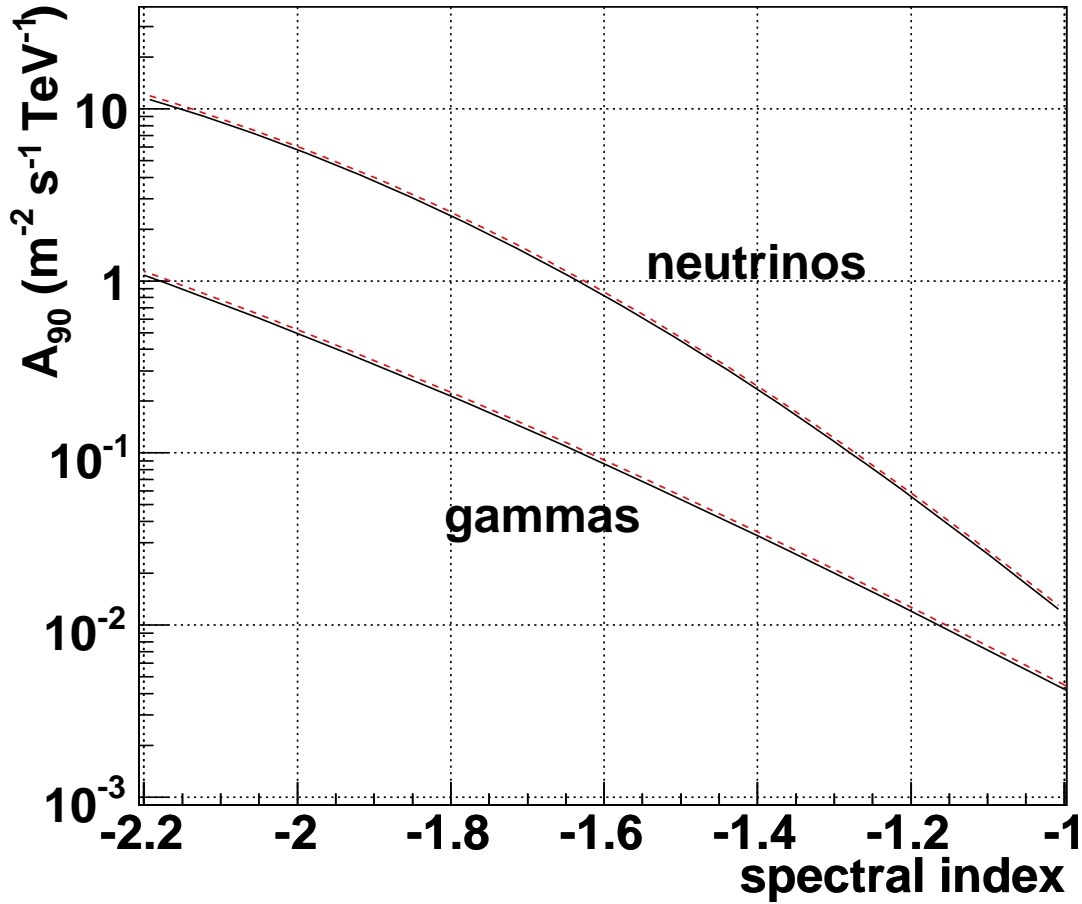} &
\end{tabular}
 \end{center}
 \caption{Sensitivity (dashed, red) and limit (solid, black) to the
 normalization constant in the flux of gammas (lower, thick line) and neutrinos
 (upper, thin line), assuming a flux $\phi(E)=A~(E/{\rm TeV})^{\gamma}$.}
 \label{fig:limits}
\end{figure}

{\it Results.---} Once the optimum search bin size of $5.8^{\circ}$
around the source was determined, we unblinded the 1.5 s data around
the burst looking for events satisfying the analysis requirements.  No
event was observed in the on-source, on-time window.  Then, we
determined the upper limits~\cite{pole} of the normalization constant
$A_{90}$ at a $CL$ of 90\% assuming a power-law energy spectrum,
\begin{equation}
\frac{dN}{dE} < A_{90} (E/\mathrm{TeV})^{\gamma}
\end{equation}

\noindent with a cut-off at $10^{5}$~TeV. These limits are shown in
figure~\ref{fig:limits} together with the sensitivity of the detector.

To give an idea of the impact of these limits on theoretical estimates
such as the $\gamma$ flux extrapolations presented in
Ref.~\cite{ourtheory}, for spectral index $-1.47$~($-2$) the limit on
the gamma flux normalization constant is $A_{90}=0.05$
($0.5$)~TeV$^{-1}$~m$^{-2}$~s$^{-1}$. The calculation of this limit
using the same energy limits as in Ref.~\cite{ourtheory} would give
3.3 (33)~TeV$^{-1}$~m$^{-2}$~s$^{-1}$, which rule out spectral indices
$\gamma\sim-1.5$, but not softer (assuming a maximum gamma energy of
500~TeV). The effect of the attenuation of the gamma flux by the
cosmic microwave background and the Galactic interstellar radiation
field has been also taken into account and has been calculated from
the results of Ref.~\cite{moskalenko}.

Since the source is above the horizon (hence there is not much column
depth for neutrinos to interact), the neutrino flux limits are an
order of magnitude worse than the TeV $\gamma$ limits, but can still
be used to constrain models.  In cases where there is large baryonic
outflow, high-energy neutrinos are produced and the baryons may make
the source partially opaque to high-energy photons.  Comparing the
extrapolations in Ref.~\cite{ourtheory}, for spectral index $-1.47$
($-2$) the limit on the $\nu_{\mu}$ flux normalization is $A_{90}=0.4$
($6.1$)~TeV$^{-1}$ m$^{-2}$ s$^{-1}$ while the model predicts
(accounting for oscillations) $1.7$ ($4.1\times10^{-4}$) in the same
units. We are thus able to exclude an extremely hard neutrino spectrum
extrapolated from the measured MeV photon flux.  On the other hand,
our limit on the high-energy neutrino fluence is still at least one
order of magnitude larger than the fluence predicted in
Ref.~\cite{Gelfand}.

{\it Conclusions.---} In summary, we have searched for TeV gammas and neutrinos associated
with the Dec. 27th giant flare from \sgr.  We demonstrate that
underground neutrino arrays such as AMANDA and IceCube can be used as
TeV $\gamma$ detectors for transient events. An analysis of AMANDA
data yields no muons coincident with the flare. We use this muon
non-observation to place stringent limits on TeV radiation from this
source.

{\it Acknowledgments.---} We acknowledge the support from the
following agencies: National Science Foundation-Office of Polar
Program, National Science Foundation-Physics Division, University of
Wisconsin Alumni Research Foundation, Department of Energy, and
National Energy Research Scientific Computing Center (supported by the
Office of Energy Research of the Department of Energy), the
NSF-supported TeraGrid system at the San Diego Supercomputer Center
(SDSC), and the National Center for Supercomputing Applications
(NCSA); Swedish Research Council, Swedish Polar Research Secretariat,
and Knut and Alice Wallenberg Foundation, Sweden; German Ministry for
Education and Research, Deutsche Forschungsgemeinschaft (DFG),
Germany; Fund for Scientific Research (FNRS-FWO), Flanders Institute
to encourage scientific and technological research in industry (IWT),
Belgian Federal Office for Scientific, Technical and Cultural affairs
(OSTC); the Netherlands Organisation for Scientific Research (NWO);
M.~Ribordy acknowledges the support of the SNF (Switzerland);
J.~D.~Zornoza acknowledges the Marie Curie OIF Program (contract
007921).

\end{document}